\documentclass[10pt]{elsart}
\usepackage{psfig}
\def\be{\begin{equation}}
\def\ee{\end{equation}}
\def\tr{{\rm Tr}}
\def\bea{\begin{eqnarray}}
\def\eea{\end{eqnarray}}
\def\tr{{\rm Tr}}
\def\ra{\rangle}
\def\la{\langle}
\newcommand{\vrho}{\varrho}
\newcommand{\ua}{\uparrow}
\newcommand{\da}{\downarrow}
\newcommand{\rmd}{{\rm d}}
\newcommand{\pbh}{{\it PBH}}
\begin{document}
\begin{frontmatter}  

\title{Prolegomena to a Non-Equilibrium Quantum Statistical Mechanics}
\author{C. Adami and N. J. Cerf}

\address{W. K. Kellogg Radiation Laboratory\\ California Institute 
           of Technology, Pasadena, California 91125, USA}

\maketitle

\begin{abstract}
We suggest that the framework of quantum information theory, which has
been developing rapidly in recent years due to intense activity in
quantum computation and quantum communication, is a
reasonable starting point to study non-equilibrium quantum statistical
phenomena. As an application, we discuss the non-equilibrium quantum
thermodynamics of black hole formation and evaporation.
\end{abstract}
\end{frontmatter}

\section{Introduction}
The classical statistical theory of thermodynamical phenomena,
due largely to Boltzmann, Maxwell, and Gibbs, is one
of the cornerstones of 20th century physics. It describes equilibrium
phenomena ranging from gas dynamics over steam engines to crystals,
while its quantum extension accurately describes radiation phenomena,
metals, and superconductivity, to name but a few examples. Nature's
tendency to move towards equilibrium following a perturbation---captured by
Boltzmann's second law---implies that most every\-day-life phenomena are
indeed taking place in an equilibrated system, for which this theory is
applicable and eminently successful. For the brief {\em transitory} periods,
however, the time during which a system {\em approaches}
equilibrium, our bag of tricks---containing the tools of statistical 
mechanics---is of little use. The canonical phenomena of this type are
relaxation or transport processes, phenomena which are usually termed
``irreversible'', and phase transitions for which the entropy is not a
constant. 

The standard approach to deal with such situations is to study the
$N$-body dynamics of the system, with a Hamiltonian that includes an
interaction term (in equilibrium statistical mechanics the Hamiltonian
is a sum of non-interacting one-body terms) and the construction of
equations that follow the $N$-particle distribution function through
time: the Boltzmann equation (see, e.g., \cite{prigogine}). 
This approach suffers from the drawback
that it can only be solved in perturbation theory, which obscures the
relation to the ``exact'' formalism of thermodynamics. In this paper,
we would like to explore the possibility that a formalism well-known from
engineering---Shannon's statistical theory of information---provides a
bridge between equilibrium and non-equilibrium statistical phenomena,
and that its quantum extension (developed primarily in support of the
recent efforts in quantum computation and communication) represents an
adequate framework to investigate certain quantum statistical phenomena that
have so far resisted a satisfying treatment.  Naturally, however, we
should not expect that the classical and quantum theory of information
provides a complete theory of all non-equilibrium phenomena. For
most dynamics with complicated time-dependent interactions and
many-body correlations, a transport-equation approach will still be
the only tractable alternative.

Standard non-equilibrium phenomena are usually termed
``irreversible'', an adjective that captures a practical aspect---a
direction of time---which, however, we know not to be
fundamental. Rather, time-reversal invariance guarantees that all dynamics can,
in principle, be reversed as long as the participating degrees of
freedom can be controlled. Even though this is clearly not always possible in
practice, it may appear as an oversight that a
practical limitation seems to be at the origin of a theorem---the
second law of thermodynamics. Indeed, as irreversibility is only
practical, so must be the second law. If we were, then, able to devise
a formalism in which the second law is replaced by a {\em conservation
  law} for entropy (and in which case the second law would appear as
a corollary) we may then be in possession of a formalism that can
quantitatively describe even the {\em approach} to equilibrium and
other non-equilibrium statistical phenomena. It is the purpose of this
paper to point out that this formalism exists in the form of the
classical theory of information, introduced by
Shannon~\cite{shannon}. Its extension to the quantum
regime (see, e.g.,~\cite{steane} and references therein)
is particularly interesting as it consistently describes quantum unitary
dynamics which dictates that the von Neumann entropy---the quantum
extension of the Shannon entropy---is a {\em constant}.  

In the next section we begin by describing the classical
statistical theory of information in physical terms (as
opposed to the more engineering-oriented approach given in most
textbooks~\cite{textbooks}). We then apply it to two classical
non-equilibrium statistical processes---meas\-ure\-ment, and
equilibration of an ideal gas---to demonstrate the use of the
formalism in physics.  In Section 3 we formulate the quantum theory
with special emphasis on those aspects that differ from the classical
theory, and discuss the EPR paradox as an illustration. We
present an application to black hole formation and
evaporation---a quintessential non-equilibrium scenario---in Section
4.  We close with conclusions and comments in Section 5. Readers
familiar with the information-theoretic approach to classical and
quantum statistical phenomena may skip directly ahead to Section 4.

\section{Classical Theory}

The intimate
relation between information theory and statistical mechanics has been
pointed out earlier by Jaynes~\cite{jaynes} in order to {\em justify} 
statistical mechanics via information theory. Here, we use information
theory to {\em extend} statistical mechanics to the non-equilibrium
regime. 

The concept of entropy was introduced by Shannon with respect to {\em
  random variables}. For a random variable $X$ that can take on values
$x_1,\cdots,x_N$ with probabilities $p_1,\cdots,p_N$ respectively, the
Shannon {\em uncertainty} (or entropy) is given by  
\be
H(X)=-\sum_{i=1}^Np_i\log p_i\;. \label{shanent}
\ee
Instead of random variables,
however, we may imagine any physical system with enumerable degrees of
freedom and enumerable states $x_i$. As is well-known and we show
  below, the Shannon entropy then represents the {\em physical} entropy
  of the system.  In fact, this concept of entropy can be
expanded to cover continuous variables, where it will suffer from the
same ambiguity (redefinition up to a constant) as standard
  thermodynamical entropy.
For the moment, let us confine ourselves to discrete degrees of
freedom and imagine that any continuous variables are {\em coarse-grained}
(either by assuming appropriate boundary conditions, or else
  artificially.) 

The relation to Boltzmann-Gibbs entropy becomes manifest if we
consider not {\em general} probability distributions
$\left\{p_i\right\}$, but an equilibrium distribution where the $p_i$
are given by the Gibbs distribution:
\be
p_i=\frac1Ze^{-E_i/kT}\;, \label{boltz}
\ee
where $E_i$ is the {\em energy} of state $x_i$, and $p_i$ then
represents the {\em probability} of $X$  to take on energy $E_i$. Note
that this probability
is normalized by the partition function $Z=\sum_ie^{-E_i/kT}$. Inserting (\ref{boltz}) into
Eq.~(\ref{shanent}) produces
\be
H = \frac{\la E\ra}{kT}+\log Z = \frac1{kT}(\la E\ra -F)\; \label{class}
\ee
and confirms that the Shannon entropy is just the standard physical entropy
in statistical mechanics and thermodynamics when rescaled by
the Boltzmann constant $k$:
\be
S = k H\;.
\ee
Above, we defined the free energy $F=-kT\log Z$ in the usual manner.
Similarly, thermodynamical averages are obtained via
\be
\la A\ra = \frac1Z\sum_{i=1}^N A_i e^{-E_i/kT}
\ee
for an observable $A$ that takes on the value $A_i$ in state $x_i$. 

Returning to random variables for a moment, imagine an additional
variable $Y$ that takes on states $y_1,\cdots,y_N$ with probabilities
$p_1'\cdots,p'_N$. We can then define the
conditional probability of finding $X$ in state $x_i$, {\em given}
  that $Y$ is in state $j$
\be
p_{i|j}=\frac{p_{ij}}{p'_j}\;,
\ee
where $p_{ij}$ is the {\em joint} probability to find $X$ in state
$x_i$ and simultaneously $Y$ in state $y_j$. 
This concept will allow us to quantify {\em correlations} between
degrees of freedom, a particularly important task in non-equilibrium
systems. Indeed, equilibrium can be {\em defined} as the state where ``all
`fast' things have happened and all the `slow' things
not''~\cite{feynman}, which implies that all non-permanent correlations
have vanished in equilibrium. 

Armed with conditional probabilities, we can define the {\em conditional
entropy} of system $X$ {\em given} that $Y$ is in, say, state $y_j$, 
i.e., the entropy of $X$ if we
are fully aware that $Y$ is in state $y_j$, or in other words, 
the {\em remaining} entropy of $X$ if 
$Y$ is held fixed in state $y_j$.  Naturally,
this is defined as 
\be H(X|y_j) = -\sum_{i}p_{i|j}\log{p_{i|j}}\;.
\ee
Also, the {\em average conditional entropy} of $X$ given $Y$ is in
{\em any} fixed state, or quite generally is {\em known}, is then
\be 
H(X|Y)=\la H(X|y_j)\ra = -\sum_{ij}p_{ij}\log p_{i|j}\;. \label{condent}
\ee
The vertical bar in the expression $H(X|Y)$ denotes the conditional
nature of the entropy, and is usually read as ``X given Y'', or ``X
knowing Y''. 

Armed with the conditional (or remaining) entropy, we can find a
measure for the amount of correlation between two systems. This is
just the ordinary entropy minus the remaining entropy if one of the
system's variables are known: the {\em shared} entropy (also called
correlation, or mutual, entropy)
\be
H(X:Y) = H(X)-H(X|Y)\;.
\ee

This is the central quantity introduced by Shannon: the mathematical
measure of {\em information}\footnote{The colon between $X$ and $Y$ is
  customarily used to indicate a shared entropy, and reminds us that
  correlation entropy is symmetric: $H(X:Y)=H(Y:X)$.}. The relation
between unconditional (also called ``marginal'') entropies such as
$H(X)$ or $H(Y)$, mutual, and conditional entropies
are best visualized by {\em Venn diagrams}. In Fig.~1, the area of
each circle represents an entropy, whereas the union of both circles
represents the joint entropy $H(XY)$.

\begin{figure}
\caption{Entropy Venn diagram for two random variables $X$ and $Y$.}
\label{fig1}
\vskip 0.5cm
\par
\centerline{\psfig{figure=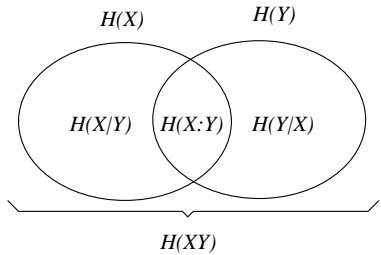,width=5cm,angle=-90}}
\vskip 0cm
\par
\end{figure}

It is straightforward to see that these quantities can be translated
into thermodynamics, by replacing the arbitrary probability
distributions by equilibrium ones. We can see immediately, however,
why they play no role in equilibrium thermodynamics. The probability
of system $X$ to take on energy $E_i$ if $Y$ has energy $E_j$ is
trivial: it is just given by $Z^{-1}e^{-E_i/kT}$ simply {\em because} $X$ and
$Y$ are in equilibrium. Thus, in equilibrium, $H(X|Y)=H(X)$, and
$H(X:Y)=0$. Away from equilibrium, conditional and mutual {\em
  thermodynamical} entropies become crucial, as we now
see. 

\subsection{Measurement}
We first treat the dynamics of classical {\em measurement}. A
measurement involves the contact between two equilibrated systems,
usually at different temperatures. The measurement device is
constructed in such a manner as to induce correlations between some of
its variables---the ``pointer''---and the measured system's degrees of
freedom (those which we desire to measure). After the initial contact
between the systems and subsequent relaxation, equilibrium is
re-established but thermodynamics seems to offer a paradox: the
entropy of the measured system appears to have been
reduced. Furthermore, this reduced entropy {\em can} be used to
perform work---in apparent violation of the second law (this puzzle is
usually termed the {\em Maxwell demon} paradox, see, e.g., \cite{demon}). 
While this dynamics is again practically irreversible, we can describe
what happens in terms of the entropies introduced above.

Before the measurement, the system (denoted by $S$) is independent of
the measurement device (denoted by $M$, see Fig.2a). They do not share
any entropy, which implies that knowledge of any one of the systems
will not allow any predictions about the other. Bringing the two
systems into contact introduces correlations, and reduces the {\em
  conditional} entropy of both $S$ and $M$. Note that before
measurement, $H(S|M)\equiv H(S)$. The amount by which the conditional
entropy is reduced is of course just the {\em acquired information},
or shared entropy $H(S:M)$ (see Fig.~2b). This shared entropy plays a
fundamental thermodynamical role: for example it can be shown that
erasing it requires the dissipation of an equal amount of
heat~\cite{landauer}. Needless to say, the marginal entropy did not
really decrease in this process, but rather {\em stayed constant}.
In contrast, the conditional entropy of $S$ is reduced, as can be seen
by inspection of the diagram in Fig.~2b,
\begin{figure}[t]
\caption{Rearrangement of entropies in the measurement process. (a)
  System $S$ and device $M$ are uncorrelated ($H(S:M)=0$). (b) Device
  and system share entropy $H(S:M)$ and the conditional entropy of
  both system and device are reduced.}
\label{fig2}
\vskip 0.5cm
\par
\centerline{\psfig{figure=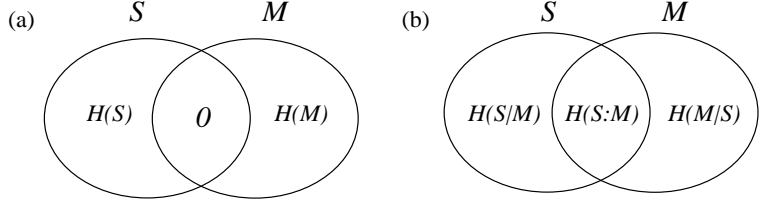,width=10cm,angle=-90}}
\vskip 0.5cm
\par
\end{figure}
\be
H(S)\longrightarrow H(S|M) = H(S) - H(S:M)\;. \label{master}
\ee
Turning Eq.~(\ref{master}) around:
\be
H(S)=H(S|M)+H(S:M)
\ee
demonstrates that non-equilibrium dynamics affects only the
distribution of $H(S)$ into either (conditional) entropy or
information, that the two however always add up to $H(S)$. 

\subsection{Equilibration}
Another example of irreversible dynamics is the notorious
``perfume bottle'' experiment, in which a diffusive substance (let's
say, an ideal gas) is allowed to escape from a small container into a
larger one. Both the initial and the final state of the system is in 
equilibrium; common wisdom however states that the entropy of the gas is {\em
increasing} during the process, reflecting the non-equilibrium
dynamics. We shall now show that this is not the case, by describing
the gas in the smaller container by a set of variables
$A_1,\cdots,A_n$, one for each molecule. The entropy $H(A_i)$ thus
represents the entropy per molecule. The entire
volume, on the other hand, is described by the {\em joint entropy}
\be
H_{\rm gas}=H(A_1\cdots A_n)\;, \label{joint}
\ee
which can be much smaller than the sum of per-particle entropies, the
standard (equilibrium) thermodynamical entropy $S_{eq}$
\be
H(A_1\cdots A_n)\ll \sum_{i=1}^n H(A_i)=S_{eq}\;.
\ee
The difference is given by the $n$-body correlation entropy
\be
H_{\rm corr}= \sum_{i=1}^n H(A_i)-H(A_1\cdots A_n) \label{corr}
\ee
which can be calculated in principle, but becomes cumbersome already
for more than three particles. 

We see that in this description the molecules after occupying the
larger volume cannot be independent of
each other, as their locations are {\em in principle} correlated (as they
all used to occupy a smaller volume, see
Fig.~3a). These correlations are not manifest in two-- or even
three-body correlations, but are complicated $n$-body correlations
which imply that their positions are not independent, but linked by
the fact that they share initial conditions. Again, this state of
affairs can be summarized by turning around Eq.~(\ref{corr})
\be
H(A_1\cdots A_n) =\sum_{i=1}^n H(A_i)-H_{\rm corr}\;.
\ee
We assume that before the molecules are allowed to escape, they 
are uncorrelated with respect to each other: $H_{\rm corr}=0$, and all
the entropy is given by the extensive sum of the per-molecule entropies.
After expansion into the larger volume, the standard entropy increases
because of the increase in available phase space, but this increase is
balanced by an increase in the correlation entropy $H_{\rm corr}$ in
such a manner that the actual joint entropy of the
gas, $H_{\rm gas}$, remains unchanged. 

Note that this description is not, strictly speaking, a {\rm redefinition} of
thermodynamical entropy. While in the standard theory entropy is an
{\em extensive}, i.e., additive quantity for uncorrelated systems, the
concept of a thermodynamical entropy in the absence of equilibrium
distributions has been formulated as the number of ways to realize a
given set of occupation numbers of states of the joint system 
(which gives rise to
(\ref{shanent}) by use of Stirling's approximation, see, e.g.,
\cite{wannier}) and is thus fundamentally {\em non-extensive}. 
Assuming the systems $A_i$ are uncorrelated reduces $H(A_1\cdots A_n)$
to the extensive sum
$\sum_{i=1}^{n}H(A_i)$, and thus to an entropy proportional to the volume
the systems inhabit. From a calculational point of view the present
formalism does not represent a great advantage in this case, as the correlation
entropy $H_{\rm corr}$ can only be obtained in special situations,
when only few-body correlations are important. 

The examples of non-equilibrium processes treated here (measurement
and equilibration) suggest that:

\begin{quote}
\em In a thermodynamical equilibrium or non-equilibrium process, the
  unconditional (joint) entropy of a closed system remains a constant.
\end{quote}

This formulation of the second law directly reflects probability
conservation (in the sense of the Liouville theorem), 
and allows a quantitative description of the amount by
which the conditional entropy is decreased in a measurement, or the
amount of per-particle entropy is increased in an equilibration process.

\begin{figure}
\caption{Diffusion of an ideal gas from a small into a larger container. (a)
The molecules with entropy $H(A_1\cdots A_n)$ occupy the smaller
  volume, and their correlation entropy is zero. (b) The molecules
  have escaped into the larger container, which increases the sum of
  the per-particle entropies and increases the correlation entropy
  commensurately such that the overall entropy remains unchanged.}
\label{fig3}
\vskip 0.5cm
\par
\centerline{\psfig{figure=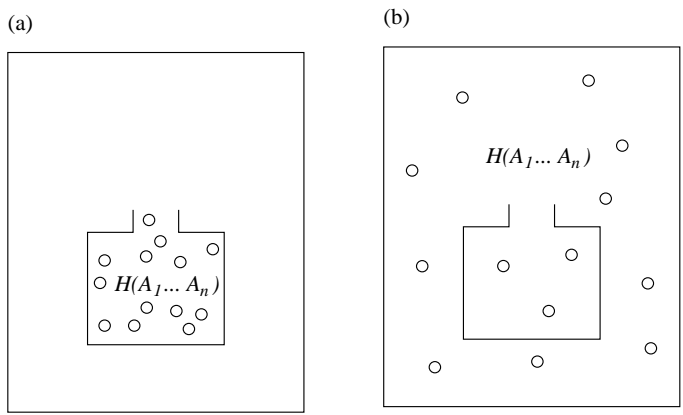,width=9cm,angle=-90}}
\vskip 0cm
\par
\end{figure}

\section{Quantum Theory}
As the classical non-equilibrium mechanics described above is founded
on the classical theory of information, its quantum 
extension is built on the quantum theory of information introduced
recently~\cite{schum,ca1,ca2}. 

\subsection{Equilibrium}
For our purposes, equilibrium quantum statistical mechanics can be
summarized in a few equations. For a system described by 
Hamiltonian\footnote{In the following, $H$ stands for the Hamiltonian,
  while entropies are denoted by the symbol $S$.}
 $H$ and partition function (we set $\beta=1/kT$ from now on)
\be
Z=\tr\, e^{-\beta H}\;,
\ee
the density matrix can be written as
\be
\vrho = \frac{e^{-\beta H}}Z \label{equilib}
\ee
while the free energy is
\be F = -\frac1\beta \log Z\;.
\ee
Accordingly,  
\be
\log \vrho = \beta F -\beta H \label{logrho}
\ee
and, defining the internal energy  $U=\tr\,\vrho H$, we obtain the 
equivalent of Eq.~(\ref{class})
\be
S = \beta (U-F)
\ee
where 
\be
S(\varrho)=-\tr\, \vrho \log \vrho\label{vnentropy}
\ee 
is the quantum entropy of the state described by the density
matrix $\varrho$, introduced by von Neumann~\cite{vn}. 
While we used equilibrium
expressions to motivate (\ref{vnentropy}), it is in fact valid even
when an equilibrium expression such as (\ref{equilib}) does not
exist. Just as the classical entropy (\ref{joint}), this entropy
remains a constant under {\em any} dynamics, reversible or
irreversible. This is in fact more obvious in the quantum
case, as the density matrix $\vrho$ is known to evolve in a unitary
manner
\be
\vrho(t)=U(t)\vrho(0)U^\dagger(t) \label{unitary}
\ee
which immediately implies, using (\ref{vnentropy}) and the cyclic
property of the trace, that 
\be
\frac{\rm d}{{\rm d}t} S(t) = 0\;.
\ee

Inserting (\ref{equilib}) into (\ref{vnentropy}) on the other
hand allows us to recover the 
Boltzmann-Gibbs-Shannon entropy (\ref{shanent}), with the
probabilities given by 
\be
p_i=\frac1Z \,e^{-\beta E_i} 
\ee
with $E_i$ the eigenvalues of $H$. In general, when considering the
diagonal elements of $\vrho$ in a basis distinct from the eigenbasis
of $H$, the von Neumann entropy is a lower bound on the
Boltzmann-Gibbs-Shannon entropy
\be
S(\vrho)\leq -\sum_ip_i\log p_i\;,
\ee
where the equality holds for density matrices $\vrho$ that are
diagonal, in which case quantum statistical mechanics is formally
identical to the classical description. Differences arise for
non-diagonal $\vrho$. The off-diagonal terms signal the presence of
quantum {\em superpositions} and the potential for {\em
  entanglement}---a form of ``super-correlation''.
As we shall see, entanglement requires a radical departure from the
classical description, and an extension of the above formalism to a 
non-equilibrium quantum statistical mechanics. 

\subsection{Non-equilibrium}
As mentioned earlier, in classical mechanics equilibrium between two
ensembles $A$ and $B$ implies that all ``fast'' degrees of freedom are
independent (no correlations) whereas the ``slow'' degrees are
considered to be static. This is usually achieved by waiting for times
larger than the relaxation time.
The situation is dramatically
different in quantum mechanics. As we shall see, entanglement
introduces a type of super-correlation that cannot be undone by
letting the system equilibrate, not even if the two systems are
separated by space-like distances.

As an example, consider the joint system $AB$ where $A$ and $B$ are
half-integral spin states with eigenstates $|\uparrow\ra$ and
$|\downarrow\ra$. It is then possible to construct a wavefunction for
the joint system $AB$ which makes it mathematically and logically
impossible to attribute a {\em state} to either $A$ or $B$ by itself: the
well-known EPR state
\be
|\Psi_{AB}\ra=\frac1{\sqrt{2}}\left(|\ua\ua\ra-|\da\da\ra\right)\;. \label{epr}
\ee
However, both $A$ and $B$ can be described by {\em reduced} density
matrices, obtained by tracing $B$ or $A$ out of 
the joint matrix $\vrho_{AB}$
\be
\vrho_{A(B)}=\tr_{B(A)}\vrho_{AB}=
\frac12\biggl(|\ua\ra\la\ua|+|\da\ra\la\da|\biggr)\;,
\ee
where $\tr_{B(A)}$ denotes the partial trace over $B(A)$. 
As these density matrices are diagonal, the quantum entropy is just
equal to the classical one
\be
S(A)=S(B)=1
\ee
if we agree to take base-2 logarithms and count entropy in
``bits''. The joint entropy $S(AB)$ on the other hand is {\em not}
equal to 2, i.e., the entropy is {\em non-extensive}. As we mentioned
earlier, this implies that correlations are present and calls for a
non-equilibrium formalism. Things are worse here.  
For this wavefunction, the quantum entropy {\em vanishes} (it is a
pure state: the only non-vanishing eigenvalue of the density matrix
$\vrho_{AB}= |\Psi_{AB}\ra \la \Psi_{AB}|$ is 1.) This well-known property of
quantum mechanically entangled systems is known as the {\em
  non-monotonicity} of quantum entropy (see, e.g., \cite{wehrl}) and
forces us to rethink the equilibrium formalism that we recapitulated
earlier. We will proceed in a manner similar to the non-equilibrium
classical mechanics of the previous section, by introducing quantum
{\em conditional} and {\em mutual} entropies. As in the classical
case, the conditional quantum
entropy then would reveal to us the entropy of a quantum system {\em
  given} we know the state of another system it is entangled with,
while the quantum mutual entropy would reflect the amount of
correlation between the systems. In contrast to the classical
situation, quantum conditional entropies can be {\em negative}, while
the mutual entropy can {\em exceed} the classically allowed limit (hence the
term super-correlation.) This formalism has turned out to be useful in
the information-theoretic analysis of quantum
measurement~\cite{ca2,meas}, as well as the description of the
non-equilibrium physics of quantum information
transmission~\cite{channel}. 

Guided by the classical case, we are tempted to define the conditional
quantum entropy of system $A$ given the state of $B$ by
\be
S(A|B)= S(AB)-S(B)\;,
\ee
i.e., the quantum entropy of the joint system minus the entropy of $B$
(as that is given). This structure then suggests an expression for the
{\em conditional amplitude matrix} $\vrho_{A|B}$, which we need to
formulate the non-equilibrium dynamics. This matrix, first introduced
in \cite{ca1}, is a well-defined Hermitian operator on the joint
Hilbert space of $A$ and $B$ (see \cite{CAG}) defined by
\be
\vrho_{A|B}=\exp[\log\vrho_{AB}-\log({\bf 1}_A\otimes\vrho_B)]
\label{condmat}
\ee
which allows us to write
\be
S(A|B) = -\tr \vrho_{AB}\log \vrho_{A|B}
\ee 
in analogy with (\ref{condent}). In contrast to the classical
conditional probability $p_{i|j}$, the conditional amplitude matrix
can have eigenvalues {\em exceeding} unity, which reflect the quantum
inseparability of the system. 

\begin{figure}[t] 
\caption{Quantum entropy Venn diagrams. (a) Definition of
joint [$S(AB)$] (the total area), marginal [$S(A)$ or $S(B)$], 
conditional [$S(A|B)$ or $S(B|A)$] and mutual [$S(A:B)]$ entropies for a 
quantum system $AB$ separated into two subsystems $A$ and $B$; (b)  
their respective  values for the EPR pair.
\label{fig4} }
\vskip 0.25cm
\centerline{\psfig{figure=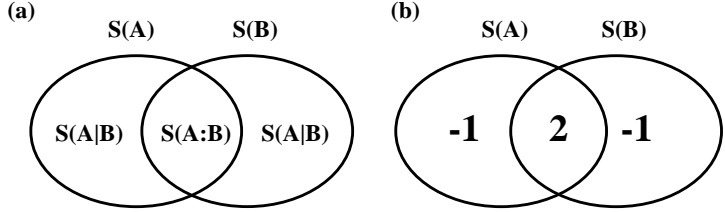,width=3.75in,angle=0}}
\vskip 0.5cm
\end{figure}

The mutual quantum entropy can be defined in an analogous manner
\be
S(A:B) = S(A)-S(A|B)
\ee
as the marginal (unconditional) quantum entropy of $A$ minus the
``remaining'' entropy $S(A|B)$. Consequently, we can extend the useful
Venn diagram technique (Fig.~1) to the quantum regime, and just
replace $H$ by $S$ (Fig.~\ref{fig4}a). 
The peculiarity of quantum superpositions such as
the EPR wavefunction Eq.~(\ref{epr}) is immediately apparent in its
Venn diagram (Fig.~\ref{fig4}b). 

More generally, a mixed state $\vrho=\sum_i p_i |i\ra\la i|$ can
always be ``purified'', i.e., written as the partial trace over a pure
state $|\psi\ra=\sum_i\sqrt(p_i)|i\ra|i\ra$ by means of the Schmidt
decomposition, while being represented by a Venn diagram such as
Fig.~\ref{fig4}b but with entries $\{-S,2S,-S\}$ instead of
$\{-1,2,-1\}$, where $S=-\sum_i p_i\log p_i$. Furthermore, the diagram
technique and the use of quantum entropies can easily be extended to
understand the quantum correlations between three systems.  An
instructive example is the description of the EPR
paradox~\cite{reality}, which we briefly summarize as it is relevant
to the discussion of black holes which follows.

\begin{figure}[t]
\caption{Measurement of EPR pair $Q_1Q_2$ by devices $A_1$ and $A_2$.
\label{meas} }
\vskip 0.25cm
\centerline{\psfig{figure=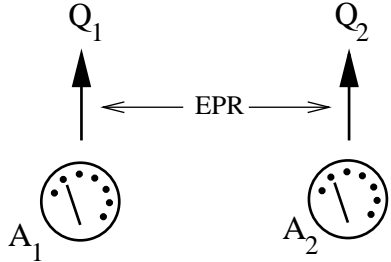,width=2.0in,angle=-90}}
\end{figure}

\begin{figure}[h]
\caption{ (a) Quantum entropy diagram for the EPR measurement of same
spin-projections: e.g., $A_1$ and $A_2$ both measure $\sigma_z$. (b)
Reduced diagram obtained by tracing over the quantum states $Q_1$ and
$Q_2$ (the dashed line surrounds degrees of freedom
traced out, i.e., averaged over) reflecting the correlation between
the measurement devices. 
\label{epr1} }
\vskip 0.5cm
\centerline{\psfig{figure=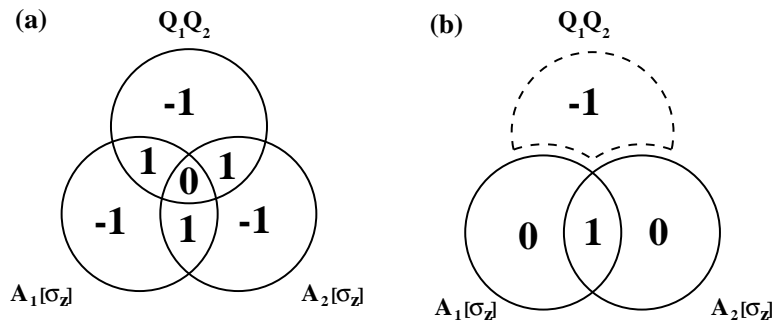,width=4.0in,angle=-90}}
\end{figure}

\begin{figure} 
\caption{ (a) Quantum entropy diagram for the EPR measurement of orthogonal 
spin-projections, e.g., $A_1$ measures $\sigma_Z$ while $A_2$ records 
$\sigma_x$. (b) Reduced diagram as above. In this case the measurement
devices show zero correlation, while entanglement persists between
quantum system and measurement devices.
\label{epr2} }
\vskip 0.5cm
\centerline{\psfig{figure=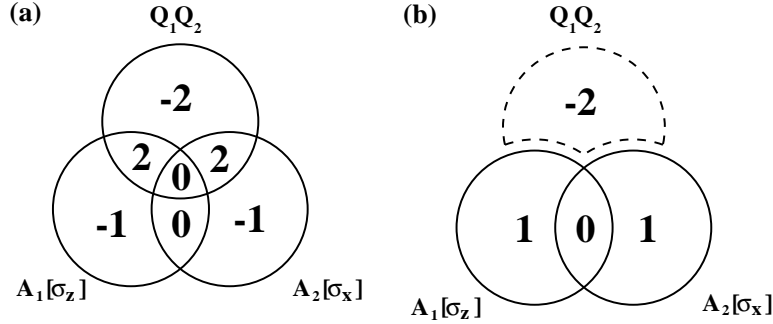,width=4.0in,angle=-90}}
\vskip 0.5cm
\end{figure}

Imagine a wavefunction such as (\ref{epr}), with the particles in
question separated by space-like distances. Imagine further that at
each of these separated locations, measurements of the spin-projection
are performed in either the $x$ or the $z$ direction. Beyond the
quantum bipartite system described by Eq.~(\ref{epr}), which we denote by
$Q_1Q_2$ in the following, we introduce Hilbert spaces for the
measurement devices, the ``ancillae'' $A_1$ and $A_2$ rigged to measure
the polarization of $Q_1$ and $Q_2$ respectively (see Fig.~\ref{meas}).
Depending on whether same (Fig.~\ref{epr1}) or orthogonal
(Fig.~\ref{epr2}) polarizations are measured at
the remote locations, the measurement devices are either correlated or
independent. However, in both cases, the entanglement between quantum
systems and measurement devices is more complicated, and even in case
the measurement devices appear uncorrelated (Fig.~\ref{epr2}b), subtle
entanglement persists.

\par

\section{Black hole Formation and Evaporation}
The discovery of Hawking radiation~\cite{hawking} appears
to have plunged quantum mechanics into a deep crisis, as it seems to
imply that the evaporation of black holes violates unitarity (for a
review, see, e.g.,~\cite{preskill}). Below, we formulate the
``information-loss'' problem in terms of the formalism described here,
and argue for a consistent description in terms of quantum
non-equilibrium thermodynamics. 

\subsection{Black hole entropy and information paradox}
Black holes have the remarkable property that they are fully described
by very few variables---a non-rotating non-charged black hole by only
one, its mass. Bekenstein~\cite{bekenstein} and Hawking~\cite{hawking}
determined that an {\em entropy} can be defined for a Schwarzschild 
black hole which
is given entirely in terms of the area $A$ inside the event horizon
\be
S_{BH}=\frac14 A\;.
\ee
This area, in turn, is just $A=4\pi R^2$ where $R$ is the radius of
the black hole given (in units where $\hbar=G=1$) by $R=2M$, so that
the black hole entropy is specified entirely in terms of the black
hole mass $M$
\be
S_{BH}=4\pi M^2\;.
\ee
While a number of reasonings lead to this expression, including the counting
of microscopic quantum states that give rise to a black hole,
Hawking~\cite{hawking1} 
pointed out that the process of thermal evaporation of a black hole
leads to an ``information paradox''. If we assume that the black hole
is formed from a quantum mechanically pure state $S=0$, the entropy of
the purely thermal
blackbody radiation left behind {\em after evaporation} should be of
the order $\sim M^2$, i.e., a pure state evolved to a
mixed one. This contradicts the unitary evolution of quantum states
Eq.~(\ref{unitary}), according to which (as we have pointed out
repeatedly) the entropy of a closed system is a constant, in this
particular case the constant zero. 

Several avenues have been proposed to escape this conclusion, and we
will focus here on the most conservative explanation, namely that
Hawking radiation is effectively {\em non-thermal} (in the sense that
quantum correlations between the radiation and the state of the black
hole exist in principle), and that a pure state {\em is} formed after
evaporation, only that it is impossible to distinguish it from
purity~\cite{page,thooft,ds}. We first note that beyond the
information paradox pointed out by Hawking, as observed by
Zurek~\cite{zurek} we also need to match the
black hole entropy $S_{BH}$ with the entropy of approximately thermal
radiation $S_{\rm rad}\sim T^3_{H}$ with black hole temperature
$T_H=(8\pi M)^{-1}$. We then proceed to propose a scenario in which
this might be achieved.

\subsection{Black hole formation from a pure state}
\begin{figure}
\caption{Venn diagrams for black hole formation. (a) Just before
  collapse. (b) After collapse. $\Sigma$ denotes the entropy of the
  proto-black-hole, while $S_{BH}$ is the Bekenstein-Hawking entropy,
  and $\Delta S$ is the entropy deficit.}
\label{fig5}
\vskip 0.5cm
\par
\centerline{\psfig{figure=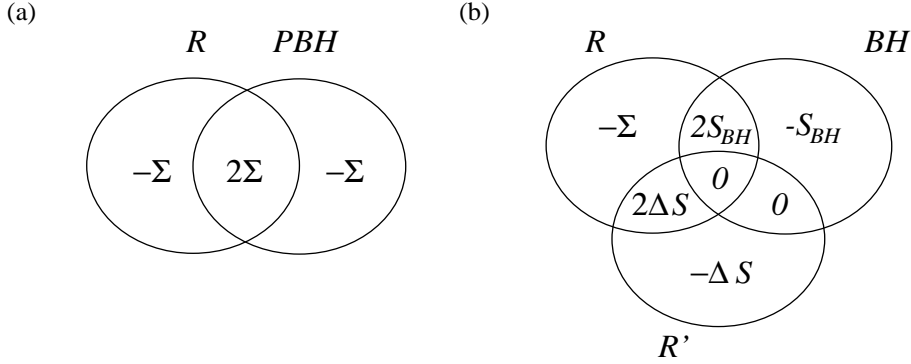,width=12cm,angle=-90}}
\vskip 0cm
\par
\end{figure}

Of course, black holes do not form by the ``collapse'' of a pure
state. Rather, we can imagine that part of a pure state
with marginal entropy $S_{\rm rad}\equiv \Sigma$ disappears behind an event
horizon. Let us divide space just before the collapse into a region
\pbh\ (the proto-black-hole) and $R$, the remainder. As the entire
system is pure ($S=0$), we know that $S_{\rm rad}=S_{PBH}$. The
entropy diagram for this situation can be constructed as described in
the previous section, and is shown in Fig.~\ref{fig5}a. 

The degrees of freedom in $R$ are
 practically inaccessible after the collapse of
the region \pbh, but we should keep in mind that they are 
{\em entangled} with \pbh\ in such a manner that the entire system,
($R,\pbh$),
is pure. In the language of quantum information theory, $R$ is a
``reference'' system that ``purifies'' \pbh. The gravitational
collapse of region \pbh\ forms an intriguing problem. While we can
assume the radiation inside it to be purely thermal, with energy
$E\sim T^4$ and corresponding entropy $\Sigma\sim4/3\,T^3$,  
the entropy of the {\em collapsed} state is $S_{BH}=4\pi M^2$, lower
than $\Sigma$. In fact, it was shown by Zurek~\cite{zurek} that the
entropy $\rmd S$ accreted by a black hole (which we can take to be of the
radiation type) is larger than the corresponding entropy increase of the black
hole itself 
\be
\rmd S \approx 4/3 \,\,\rmd S_{BH}\;,
\ee
and the same mismatch occurs in the evaporation process.

In statistical physics this is not an alarming state of affairs, but
rather is the usual scenario in a non-equilibrium phase
transition. Here, we shall mask our ignorance about the dynamics which
produces the black hole out of radiation by assigning a new {\em
  phase} to the black hole matter, and discuss the process in which
the radiation with entropy $\Sigma$ {\em condenses} to a phase with
entropy $S_{BH}$. 
  
During the condensation from the proto-black-hole state to the
black-hole ({\it BH}) state, excess
entropy $\Delta S$ has to be radiated away ($T_{H}\Delta S$ is the
equivalent of the latent heat in a first-order phase transition) .  While we
cannot offer a detailed picture of this transition, we assume that
this radiation is emitted just outside the forming horizon, and
represents the bremsstrahlung of the accelerated particles
accreting on the black hole.  This gives rise, then, to the system
depicted in Fig.~4b, where the bremsstrahlung $R'$ is entangled
with both $R$ and the black hole {\it BH}, with marginal entropy
$S(R')=\Delta S
= \Sigma - S_{BH}$.  During the phase transition,
the entropy of the \pbh\ system remains constant, but is distributed
over the joint system ({\it BH},$R'$):
\be
\Sigma =S(PBH)=S(R',BH)=S(BH)+S(R'|BH)=S_{BH}+\Delta S\;.
\ee
The ``missing'' entropy $\Delta S$ therefore is contained in radiation
$R'$ emitted during the collapse. 

This scenario, which is the time-reverse of the evaporation
process considered next, naturally leads to a radiation field $R'$
that is causally uncoupled from the black hole, as
$S(BH:R')=0$. Tracing over the ``reference'' field $R$ leads to the
trivial entropy diagram diagram $\{S_{BH},0,\Delta S\}$. We need to
keep in mind, however, that just as in the EPR situation described
previously, the wavefunctions of $R'$ and the black hole are linked
via entanglement with the quantum degrees of freedom $R$.

\subsection{Evaporation of black holes}
The processes of black hole formation and evaporation can be
considered time-reverse images of each other.
Evaporation of black holes occurs through the formation of
virtual particle--anti-particle pairs of energy $2dE$ close to the
horizon due to quantum mechanical tunneling in the
strong gravitational field. If one of the members of the pair
disappears behind the horizon while the other manages to escape, the
escaping particle appears to have a black-body spectrum with
temperature $T_{H}$, while the energy of the black hole is reduced by
$dE$. The paradox occurring here thus appears to be the same as the one
encountered in the condensation process. How does the radiation pick
up the extra entropy? In terms of quantum information theory, the creation of a
particle--anti-particle pair is akin to
the creation of an EPR state with vanishing entropy,
described by the entropy diagram in Fig.~\ref{fig4}b. However,
just as in standard first-order ``evaporation'' transitions, the
black hole has to provide in addition the latent heat for ``decondensation'',
i.e., the energy to create
the entropy $\Delta S$. Thus, a pair created with
$2\rmd E$ and temperature
$T_{H}$ will not reduce the black hole mass by an amount $\d E$, but by
\be
\Delta E=\rmd E-T_{H}\Delta S\;,
\ee
which restores the entropy and energy balance.
The entropy of the escaping particle is $\rmd S\sim T_{H}^3$
while at the same time the entropy of the black hole is reduced by
\be
\rmd S_{BH}=4\pi\left(M^2-(M-\Delta E)^2\right) =
\frac{\rmd E}{T_H}-\Delta S\;. 
\ee

Arguments have been raised (see the reviews~\cite{preskill} and in
particular~\cite{susskind}) that seem to imply that information stored
in correlations and entanglement between the black hole and its
surrounding radiation field cannot be retrieved, even in principle.
These arguments rest on the assumption that the (low-energy) quantum
fields live in a Hilbert space that is of the product form ${\mathbf
  H}_{\rm in}\otimes{\mathbf H}_{\rm out}$, and an application of the
quantum no-cloning theorem. While the fields do live in a product
Hilbert space, the wavefunction of an EPR pair
created at the event horizon of the black hole indirectly becomes
entangled with the hole the moment one of the particles crosses the horizon
(even though the quantum fields are separated by space-like distances)
and the combined quantum state becomes inseparable. This situation is
not unlike the scenario we noted in the formation of the black hole,
where the accreted particle and the radiation it emits when tumbling
into the black hole can be considered an entangled, EPR-type state
(albeit with real rather than virtual energy). Just as in that case
the radiation $R'$ shared no entropy with the black hole, neither does
the Hawking radiation, while still being entangled with it.  Thus, the
Hawking radiation carries ``information'' about the inside of the hole
in the same manner as the measurement of EPR partners separated by
space-like distances reveals correlations in measurement devices that
are at space-like distances.  
Yet, a fundamental problem remains
that is unlikely to be solved within the present formalism. The
Hawking radiation---while emitted in a unitary manner and while
information loss certainly does not take place---remains causally
uncorrelated to the black hole as long as the horizon separates the
black hole entropy from the radiation field. In a sense, we have to
wait until the last moment---the disappearance of the black hole---for
the entropy balance to be restored. This appears to put a severe
strain on current black hole models, as it is hard to imagine that
this much entropy can be stored in an ever-shrinking black hole. This
problem is likely due to our incomplete understanding of late-stage
black holes, rather than a problem intrinsic to quantum mechanics.

An alternative solution would present itself if the Bekenstein-Hawking
entropy could be understood in terms of a {\em conditional} entropy.
In that case, entropy flow from the black hole to the outside via the
formation of virtual pairs is understood easily, as the member of the
pair that crosses the horizon not only has negative energy but also
negative conditional entropy (see Fig.~\ref{fig4}b). As a
conditional entropy can become as negative as the marginal entropy of
the system it is a part of, we can circumvent the argument that ``the
black hole cannot store the information until the end because it runs
out of quantum states'', because the radiation could ``borrow'' as
much entropy as necessary from the black hole
until the horizon has
disappeared.  Within the present framework, there appears to be no
physical picture which would suggest that the Bekenstein-Hawking
entropy is in fact conditional. It is not inconceivable, however, that
a quantum statistical information theory extended to curved space-time
would reveal such a state of affairs.

\section{Conclusions}
We have used a formalism developed in the exploration of quantum
com\-pu\-ters---quantum information theory---to describe quantum processes away
from thermodynamical equilibrium, such as the formation and
evaporation of black holes. The formalism emphasizes the {\em
  conservation} of entropy, and is particularly useful in situations
where entropy is distributed over two or three systems. We emphasize
that great care is needed in using the concepts of entropy and
information consistently: information, for example, can {\em never} be
``stored'' in one system (e.g., a black hole). Rather, information is
a measure of correlation {\em between} two systems, which implies that
information is {\em always} stored in correlations. The analysis of
information storage in black hole formation and evaporation presented
here is a simple application of these rules to a scenario in which
black holes are considered special states of matter with an equation
of state different from that of radiation (or usual matter).
Transitions between those states occur continuously as the specific
heat of black hole matter is negative~\cite{hawking}.  As a
consequence, radiation and black hole matter are unstable at any time,
and transitions must occur as long as matter of either kind is
present. Yet, a consistent formulation of the correlations between
radiation and matter shows that entropy is not created during the
process, and consequently that information is conserved. Still, the mechanism
by which the pure state is restored in the last stages of black hole
evaporation may require deeper insights into quantum gravitational
dynamics, and possibly an extension of information theory to curved
space-time. 

\noindent{\bf Acknowledgments}

\noindent We are indebted to H. A. Bethe for many useful discussions,
in particular for suggesting to us to address the
impact of negative entropies on quantum statistical mechanics. This
work was supported in part by NSF Grants PHY 94-12818 and PHY
94-20470, and by a grant from DARPA/ARO through the QUIC Program
(\#DAAH04-96-1-3086).  N.J.C.  is Collaborateur Scientifique of the
Belgian National Fund for Scientific Research.

\newpage
\end{document}